\documentclass[conference,a4paper]{APSIPA2021}
\usepackage{multirow}
\usepackage{graphicx}

\usepackage{amsmath}
\usepackage[psamsfonts]{amssymb}
\usepackage{amsxtra}
\usepackage{threeparttable}

\begin{document}

\pagestyle{empty}
\onecolumn
\noindent {\large \copyright ~2021 IEEE.  Personal use of this material is permitted.  Permission from IEEE must be obtained for all other uses, in any current or future media, including reprinting/republishing this material for advertising or promotional purposes, creating new collective works, for resale or redistribution to servers or lists, or reuse of any copyrighted component of this work in other works. \\~\\

\noindent This is a peer-reviewed and accepted version of the following in press document.} \\~\\

\noindent {\large N. McCallan, S. Davidson, K. Y. Ng, P. Biglarbeigi, D. Finlay, B. L. Lan, and J. McLaughlin. ``Seizure Classification of EEG based on Wavelet Signal Denoising Using a Novel Channel Selection Algorithm,'' {\it 13th Asia Pacific Signal and Information Processing Association Annual Summit and Conference (APSIPA ASC) (In Press)}, 2021.}

\twocolumn
\newpage

\title{Seizure Classification of EEG based on Wavelet Signal Denoising Using a Novel Channel Selection Algorithm}

\author{%
\authorblockN{%
Niamh McCallan\authorrefmark{1}\authorrefmark{3},
Scot Davidson\authorrefmark{1},
Kok Yew Ng\authorrefmark{1}\authorrefmark{2},
Pardis Biglarbeigi\authorrefmark{1},
Dewar Finlay\authorrefmark{1},
Boon Leong Lan\authorrefmark{2}, and \\
James McLaughlin\authorrefmark{1}
}
\authorblockA{%
\authorrefmark{1}NIBEC, Ulster University, Jordanstown Campus, Shore Road, Newtownabbey BT37 0QB, UK \\ } 
\authorblockA{%
\authorrefmark{2}Electrical and Computer Systems Engineering, School of Engineering, Monash University, Malaysia \\ }
\authorblockA{
\authorrefmark{3}Corresponding author. Email: mccallan-n2@ulster.ac.uk}
}

\maketitle
\thispagestyle{empty}

\begin{abstract}
  Epilepsy is a disorder of the nervous system that can affect people of any age group. With roughly 50 million people worldwide diagnosed with the disorder, it is one of the most common neurological disorders. The EEG is an indispensable tool for diagnosis of epileptic seizures in an ideal case, as brain waves from an epileptic person will present distinct abnormalities. However, in real world situations there will often be biological and electrical noise interference, as well as the issue of a multi-channel signal, which introduce a great challenge for seizure detection. 
  For this study, the Temple University Hospital (TUH) EEG Seizure Corpus dataset was used. This paper proposes a novel channel selection method which isolates different frequency ranges within five channels. This is based upon the frequencies at which normal brain waveforms exhibit. A one second window was selected, with a 0.5 second overlap. Wavelet signal denoising was performed using Daubechies 4 wavelet decomposition, thresholding was applied using minimax soft thresholding criteria. Filter banking was used to localise frequency ranges from five specific channels. Statistical features were then derived from the outputs. After performing bagged tree classification using 500 learners, a test accuracy of 0.82 was achieved. 
  
\end{abstract}

\section{Introduction}

\begin{figure}[h]
\begin{center}
\includegraphics[width=7cm]{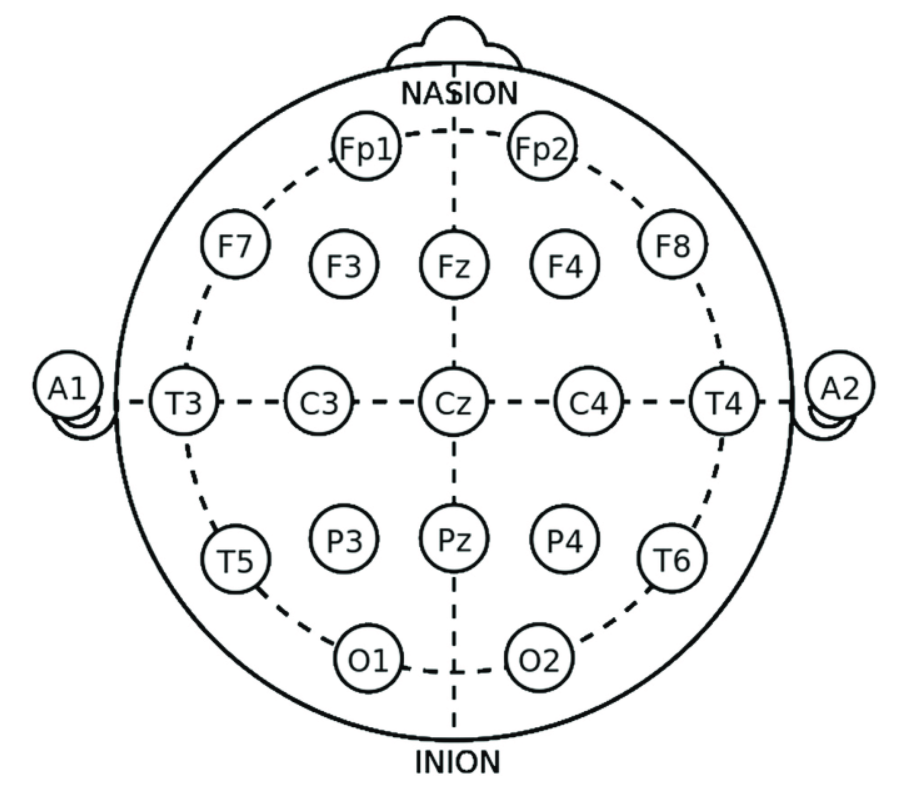}
\end{center}
\caption{EEG Electrode Configuration: International 10-20 System \cite{10_20}}
\vspace*{-3pt}
\label{fig:10_20}
\end{figure}

Electroencephalogram (EEG) is widely used in different clinical settings, with the purpose of seizure detection and classification being the most abundant \cite{eeg}. EEGs are used to measure this electrical activity by means of placing many electrodes either on the exterior of the brain using a brain cap or via intracranial electrodes \cite{eeg_prop}. Epilepsy is characterised by recurrent, unpredictable and unprovoked seizures. People with epilepsy are known to have an increased risk of injury, unemployment, death, depression, anxiety, and other psychiatric and psychological issues \cite{stig_ep}. Seizures are propagated when many neurons are synchronously excited, causing a wave of electrical activity in the brain \cite{epilepsy}. There are many different orientations that can be used for the placement of the electrodes, however, the most common method is the International 10-20 System as seen in Figure \ref{fig:10_20}. This is where 21 electrodes are evenly spaced across the scalp, with distances between each electrode equal to 10\% or 20\% of the total distance between nasion (front) and inion (back) \cite{10_20_sys}. 
This mapping is required because seizures can occur in a localised area (focal seizures) or more generally (generalised seizures). Focal seizures affect only one hemisphere of the brain and can be distinguished by whether or not awareness is retained. Generalised seizures affect the majority, if not all, of the brain and can occur without provocation \cite{roy2019machine}. It is also typical for patients experiencing generalised seizures to lose consciousness or have uncontrolled muscle spasms. It is worth noting that many people experience only one type of seizure. It is, however, typical for others to have various types of seizures. Additionally, the type of seizures that a person has may alter over time \cite{epilepsy}.

A board certified EEG interpreter must traditionally examine patients and undertake manual EEG signal analysis for diagnosis, which is expensive and time consuming. This can also be exceedingly tiresome and place a significant physical and mental strain on physicians, as EEG recordings typically span several hours, with many patients being watched overnight or even for several days \cite{saab2020weak}.
A detailed history from the patient and observers is required for an appropriate clinical diagnosis, which can be harmed by inaccurate and inadequate patient and witness histories. Recent research has revealed that even experienced neurologists have difficulty distinguishing between focal and generalised epilepsy \cite{bilinear}.
According to the World Health Organisation (WHO), if provided with appropriate diagnosis and medication, up to 70\% of people with epilepsy could avoid seizure episodes \cite{who}. As a result, substantial effort and research has gone into developing and implementing adequate seizure detection and classification algorithms to alleviate the clinical burden of manual EEG analysis \cite{sriraam2018automated}.



The International League Against Epilepsy (ILAE) defines artefact as a physiological potential difference in an EEG recording caused by something other than the brain, such as eye movement, muscle movement, or muscular contractions referred to as biological artefact. Additionally, recordings may be altered as a result of ambient electrical noise, and instrument distortion, or malfunction, referred to as technical artefact \cite{glossary}.

Early seizure detection methods rely on a variety of non-specific patient algorithms. More recently, research have focused on patient-specific algorithms to detect seizures with most findings obtaining accuracy ranging from 0.83--1 \cite{acharya2018automated,acharya2013automated}. Despite the excellent accuracy of these methods, the majority of seizure detection research studies have used the same dataset from the Department of Epileptology, University of Bonn. This dataset contains EEG recordings from 10 participants (Five without epilepsy and five with epilepsy) throughout a 23.5-second period \cite{acharya2013automated}. Therefore, it is a limited dataset to be used for seizure detection. There has been much research carried out using the TUH Seizure Corpus dataset, with Lui et al \cite{bilinear} achieving a F1 score of  0.97 and Roy et al \cite{roy2019machine} reaching 0.91. 
The purpose of this research is to concentrate on the topic of seizure detection and classification, using a large amount of annotated data. We provide a novel channel selection strategy that outperforms established methods. 
Additionally, our study demonstrates the feasibility of ensemble learning approaches over typical classification systems. 

This paper is organized as follows: Section II introduces the dataset, how signals are pre-processed alongside the novel channel selection algorithm used, feature extraction, selection and classification. It also describes the performance assessment used for this study. Section III presents the results and discussions gathered from this investigation. Section IV walks through a general conclusion of this study.


\section{Method}
This research considers three separate scenarios for seizure detection; (1) binary classification of seizure and nonseizure periods, (2) focal-generalised-nonseizure classification, and (3) multi-classification using nonseizure periods, simple-partial, complex-partial, myoclonic, absence, tonic, and tonic-clonic seizures.
\subsection{Data Acquisition}
Only a small amount of datasets are available online, free and easily accessible such as the University of Bonn Dataset \cite{bonn}, CHB-MIT \cite{shoeb} and TUH-EEG \cite{obeid2016temple}. With over 30,000 clinical EEG recordings collected over 18 years, starting in 2002 and currently ongoing, TUH Seizure Corpus has the largest publicly available dataset of EEG recordings. This dataset can be utilised for both academic and commercial purposes \cite{shah2018temple}. The reports comprise unstructured language that includes information on the patients' medical history, medications, and clinical evaluation. Based on the neurologists' report and careful study of the signal, the annotation team was able to classify the types of seizures. The data includes sessions from outpatient care, the ICU, EMU, ER, and a variety of other hospital settings. All data contains multi-channel signals which can range from 20--128 channels. A 16-bit A/D converter was used to digitise the data. The samples have a frequency range of 250--1024 Hz. More than 10 different electrode combinations and more than 40 channel configurations are included in the corpus \cite{obeid2016temple}. 

For this study, TUH Seizure Corpus dataset v1.5.1 have been utilised. Only files containing six different seizure types namely: SPSZ, CPSZ, TNSZ,TCSZ, MYSZ and ABSZ have been adopted for this investigation as per Table \ref{tab:tuh_count}. 
Our training set consists of 80\% of each seizure type and the remaining 20\% is used for testing.   
\begin{table}[t]
\centering
\caption{TUH Seizure type count}
\label{tab:tuh_count}
\begin{tabular}{cccc}
\hline
\textbf{Seizure Type} & \textbf{Total Count} \\
\hline
Simple-Partial Seizures (SPSZ)   & 8     \\
Complex-Partial Seizures (CPSZ)  & 162   \\
Tonic Seizures (TNSZ)         & 28    \\
Tonic-Clonic Seizures (TCSZ)     & 29    \\
Myoclonic Seizures (MYSZ)      & 3     \\
Absence Seizures (ABSZ)        & 20    \\ \hline
\end{tabular}
\end{table}

\subsection{Pre-processing}
Since TUH Seizure Corpus has data collected ranging from 250--1024 Hz, all signals were re-sampled to 256 Hz to ensure uniformity \cite{bilinear}. 
Only channels with EEG information were selected for further analysis. The first second of every signal was removed as this beginning segment often contains much noise. Low level frequency range, associated  with respiratory artefact, and high level frequency information is often removed to limit the bandwidth, and noise of the signal. A first order band-pass Infinite Impulse Response (IIR) filter from 0.1--80 Hz was performed on signals, followed by a 60 Hz notch filter used to remove power-line interference.  This is most typically experienced as a result of a minor problem with disconnected electrodes, which involves immediate re-connection. Without the notch filter, the signal interference would likely lead to poor tracing quality \cite{artefact}. Subsequently the signals were normalised so the range is in the interval [0,1]. This technique of signal normalisation is to scale EEG signals to the same level \cite{normalisation}. Data is segmented into one second epochs with half a second overlap. Findings of adjusting the window size clearly reveal that reducing the window increases the likelihood of seizure detection. Furthermore, studies demonstrate that by reducing the window size, predictive models can detect some peaks prior to the seizure onset. As a result, the smaller the window, the better the chance of predicting a seizure \cite{window_size}. A Kaiser window was applied to the signals with a window length of one second.

\subsection{Channel Selection}
\begin{figure}[t]
\begin{center}
\includegraphics[width=7cm]{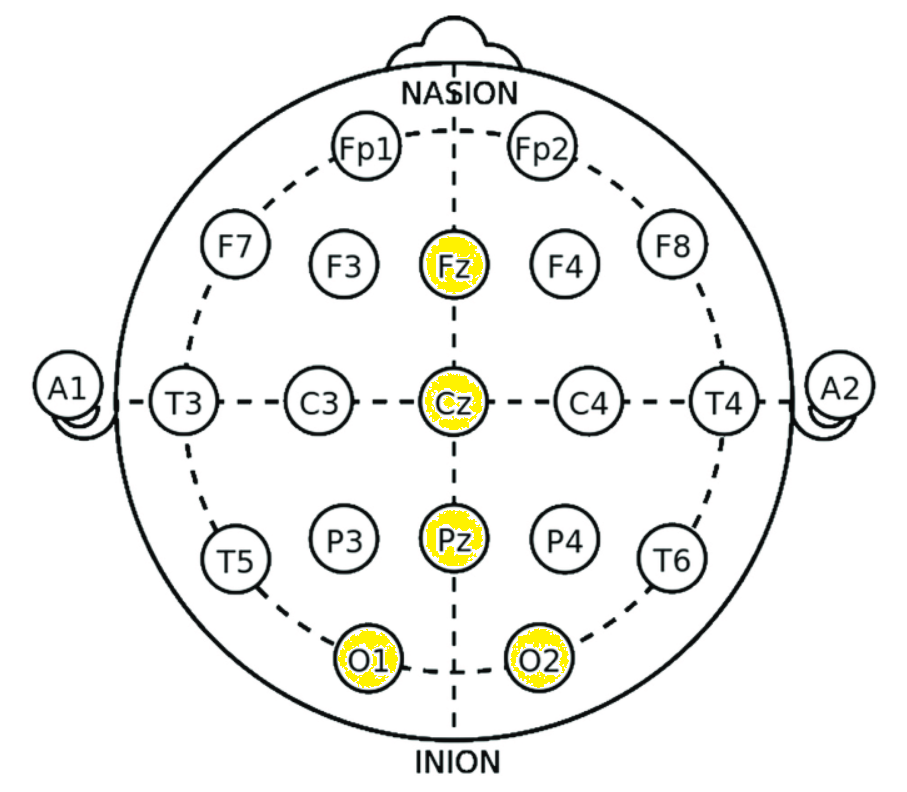}
\end{center}
\caption{EEG Electrode Configuration: International 10-20 System with selected Channel Selection.}
\vspace*{-3pt}
\label{fig:ch_10_20}
\end{figure}

A novel channel selection algorithm has been created, which maps the areas of the brain in which normal EEG waveforms in their specific frequency range should be found. The novelty of this channel selection method is that it is purely based on the regions of the brain that typical brain frequencies are dominant. This allows for further feature extraction to isolate the frequency ranges mentioned in Table \ref{table:dec_tree} and discover if a pattern can be observed to differentiate nonseizure information against seizures. 
In Nayak et al. \cite{nayak}, it has been stated that delta rhythm is prominent in the frontocentral head region. Due to light sleepiness, theta is most dominant in the frontocentral head regions and slowly migrates backward, replacing the alpha rhythm. For this reason, the frontal channel was selected as per Table \ref{table:dec_tree}. In normal waking EEG recordings in the occipital head area, the posterior dominant alpha rhythm is typically present, hence the occipital channel was selected for feature extraction. Mu rhythm is a form of alpha rhythm that manifests itself in the central head regions and has an arch-like morphology. The frontocentral head areas are where sigma waves are most noticeable. In normal adults and children, the beta rhythm is the most common rhythm. It is most noticeable in the frontal and central head regions, and it gradually fades as it moves backward, therefore, the frontal and posterior channels were selected as per Table \ref{table:dec_tree}. Attempts to locate the Gamma rhythm have initiated a lot of research around the world, although no specific localisation has been discovered and it has been attributed to different areas in the brain. In this study, channels selected include: Central (CZ), Frontal (FZ), Posterior (PZ), Occipital (O1 and O2), as seen in Figure \ref{fig:ch_10_20}.  

\begin{table}[b]
\begin{center}
\caption{Channel Selection Criteria} \label{table:dec_tree}
\begin{tabular}{ccc}
\hline
Frequency band & Frequency Range/ Hz & Electrode Positioning \\ \hline
Delta $\delta$          & 0.1--4              & FZ                    \\
Theta $\theta$          & 4--8                & FZ                    \\
Alpha $\alpha$         & 8--13               & O1, O2                \\
Beta $\beta$           & 14--30              & FZ, PZ                \\
Gamma $\gamma$          & 30--80              & CZ                    \\
Mu $\mu$             & 7--11               & CZ                    \\
Slow Sigma $\sigma $    & 12--14              & FZ                    \\
Fast Sigma $\sigma $    & 14--16              & FZ                    \\ \hline
\end{tabular}
\end{center}
\end{table}

Principle Component Analysis (PCA) and Independent Component Analysis (ICA) have been used in this study to compare the viability of our novel channel selection method as they are some of the most popular methods used for dimenionality reduction. PCA is an unsupervised method for mapping a dataset to specified feature vectors. It works by converting a high-dimensional dataset, such as multichannel EEG signals, into a low-dimensional orthogonal feature subspace, where each of the principal components is known. The variation of these principle components is organised in order of magnitude, with the first principle component having the most variance and the variance decreasing by an order of magnitude. This will limit the degrees of freedom as well as the complexities of space and time. The goal is to represent data in a space that accurately depicts variance in terms of sum-squared error. ICA is similar to PCA, but each signal is assumed to be a set of mutually independent signals. The multidimensional data is split into feature vectors that are statistically independent. \cite{subasi}.

\subsection{Discrete Wavelet Denoising}
Wavelet transform (WT) is a common approach for noise removal. Morley, a French researcher who concentrated their work on seismic data analysis, began research based on the idea of wavelet transforms in the early 1980s. Farge et al \cite{farge} contains a comprehensive description of many types of wavelet analysis methods, such as Continuous Wavelet Transforms (CWT)s and Discrete Wavelet Transforms (DWT)s. WT methods have been utilised by many researchers to reduce ocular artefacts (OA) such as in Al-Qazzaz et al \cite{dwt}, Zikov et al \cite{dwt_2}, and Krishnaveni et al \cite{dwt_3}. Wavelet transforms can provide high frequency resolution at low frequencies while also providing high time resolution at high frequencies. The DWT of a signal $x[n]$ is composed of approximation coefficients, $W\phi[j_{0},k]$, and detail coefficients, $W\varphi[j_{0},k]$.

\begin{figure}[h]
\begin{center}
\includegraphics[width=8cm,trim={0cm 0cm 0cm 0cm},clip]{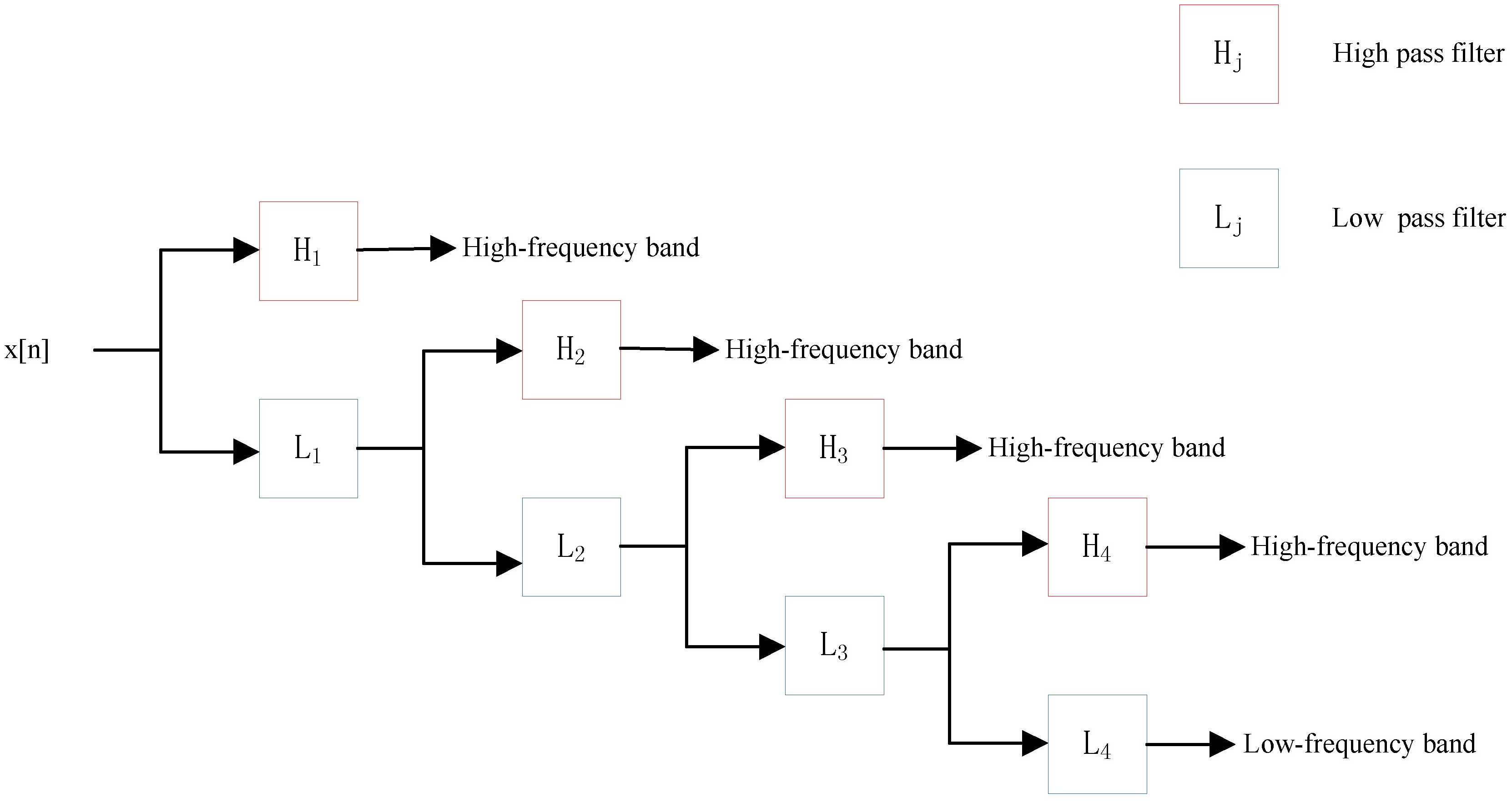}
\end{center}
\caption{A four-level DWT \cite{dwt}.}
\vspace*{-3pt}
\label{fig:dwt}
\end{figure}

The signal's approximation coefficients represent the low-frequency components derived from the original signal's low-pass filter, while the detail coefficients are obtained by passing the signal through a high-pass filter at a higher level. To compute the detail and approximation coefficients at a lower level, the signal is down-sampled by two. For a multi-level decomposition, this tree structure is repeated as seen in Figure \ref{fig:dwt}. 
For the objectives of noise reduction of the EEG data, a 4th-level wavelet decomposition using the ‘db4' Daubechies wavelet as the mother wavelet was used. With the noise estimate being level dependent, Minimax soft thresholding was adopted. Minimax is another global thresholding method developed by Donoho and Johnstone \cite{minimax}. This criteria is based on minimax principle that is used in statistics.

\subsection{Filter Banking}
From the five channels selected, 10 features where extracted by means of filter banking based on the criteria depicted in Table \ref{table:dec_tree}. A first order bandpass filter was used to split the various frequency bands. 


\subsection{Feature Selection and Classification}
Following feature extraction, statistical analysis of each one-second epoch of all 10 channels is undertaken to further reduce the dataset's dimensions. The statistics utilised include; maximum, minimum, root-mean-square (RMS), variance, standard deviation, log energy, normalised entropy, and maximum frequency. Patient age is also used as a feature, due to its general importance in seizure diagnosis \cite{epilepsy}. The equations of these statistics are as follows:
\subsubsection{Root-mean-square}
The root-mean-square level of a vector x is defined as:
\begin{equation}
    x_{RMS} = \sqrt{\frac{1}{N}\sum_{n=1}^{N}{|x_n|^2}}.
\end{equation}
with the summation taking place along the chosen dimension.

\subsubsection{Variance}
For a random variable vector $A$ made up of $N$ scalar observations, the variance is:
\begin{equation}
    V = \frac{1}{N-1} \sum_{i=1}^{N}|A_i-\mu|^2 \\
\end{equation}
where $\mu$ is the mean of $A$,
\begin{equation}
    \mu = \frac{1}{N} \sum_{i=1}^{N}A_i.
\end{equation}
Some definitions of variance use a normalisation factor of $N$ instead of $N-1$, which can be set to w as $1$. In either scenario, the typical normalisation factor $N$ is assumed for the mean.

\subsubsection{Standard Deviation}
\begin{equation}
    S = \sqrt{\frac{1}{N-1} \sum_{i=1}^{N}|A_i-\mu|^2}
\end{equation}
The standard deviation is the square root of the variance. 

\subsubsection{Log Energy}
Log energy is defined as:
\begin{equation}
\begin{split}
    LE(s_i) = log(s_i^2) \\
    LE(s) = \sum{i} log(s_i^2)
\end{split}
\end{equation}
where the convention $log(0) = 0$ is assumed. $s$ is the signal and $s_i^2$ the coefficients of $s$ in an orthonormal basis.

\subsubsection{Normalised Entropy}
The concentration in $l^p$ norm entropy with $1 \leq p$, where $p$ is the power equal to 1.1. The equation for normalised entropy is:
\begin{equation}
\begin{split}
    NE(s_i) = |s_i|^p \\ 
    NE(s) = \sum{i} |s_i|^p = ||s||_{p}^{p}
\end{split}
\end{equation}
where $s$ is the signal and $s_i^2$ the coefficients of $s$ in an orthonormal basis.

\subsubsection{Maximum Frequency}
The maximum frequency is achieved firstly by getting the Fast Fourier Tranform (FFT) of each epoch, the equation is as follows:
\begin{equation}
    Y(k) = \sum_{j=1}^{n} X(j)W_n (j-1)(k-1) \\
\end{equation}
where,
\begin{equation}
    W_n = e^{(-2\pi i)/n}
\end{equation}
After which the absolute values are achieved, using the following equation:
\begin{equation}
    abs = \sqrt{Imag^2 + Real^2}
\end{equation}

Before classification of this data could begin, some post-processing was required as the dataset was completely unbalanced, with the large majority of labels being nonseizure periods. Therefore, an algorithm was implemented to balance out this dataset, where the total count of seizure and nonseizure periods were tallied. Any addition labels beyond the mean of nonseizure periods where removed at random. 

Bagged Trees, otherwise known as Bootstrap is a prominent ensemble machine learning method that has previously demonstrated its efficiency in a variety of real-world categorisation problems. Bagged tree was first developed by Breiman in 1996 \cite{breiman}. It teaches a set of classifiers how to classify a new object \cite{bagged}. Bagging is a technique for combining classifiers to achieve higher accuracy than a single classifier. Using bootstrap re-sampling, the ensemble bagged tree (EBT) classifier separates the training data into subsets. Each subset is used as training data to build each decision tree. The bootstrapping number determines the number of decision trees that are built. The outputs of the decision trees are then used in the majority of voting stage. This model generates an ensemble of simple decision trees \cite{bagged_2}. 

For this study, bagged tree classification was used, with 500 learning cycles, k-fold cross validation was then performed using 10 sub-samples. The k-fold cross-validation algorithm divides all samples into k sub-samples at random. A sub-sample is validated using k sub-samples, and the linked classifier is tested using the remaining k-1 sub-samples. This method is carried out k times in total. For verification, each sub-sample is utilised only once. The average of k outcomes is then used to calculate a single result. As a result, all samples collected by randomly repeated sub-sampling can be used for both training and validation \cite{bagged_2}.

The k-nearest neighbour classifier has been used in this study as a comparison classifier against bagged tree classification. It is a simple, nonparametric, nonlinear classifier. For large training sets, this strategy is very effective. It is based on the training and test sets' similarity measures.
The training sets create the n-dimensional pattern space, and each set represents a point in n-dimensional space. Based on neighbouring k training data sets, a test/unknown data set is allocated to the class. Eq. (10) is used to calculate the ‘nearness' of the data sets.
\begin{equation} 
    ED = \sqrt {\sum \nolimits _{i=1}^{n} {\left ({ Y_{1i}-Y_{2i} }\right )^{2}}}
\end{equation}
where,
\begin{equation} 
{ Y}_{1i} = (y_{11},y_{12}\ldots y_{1n}) \text { and }Y_{2i}= (y_{21},y_{22}\ldots y_{2n}) 
\end{equation}
Before doing the computation on $ED$, the values of each attribute can be normalised. The classifier generally uses a majority vote from the k-nearest neighbours instead of using the single closest data set. The value of k, the number of neighbours with the lowest error rate, has been set to eight  \cite{knn}. The distance was set to cityblock, in which the distance between two points in a fixed Cartesian coordinate system is measured \cite{city} as per Eq. (\ref{eq:city}).
\begin{equation}
    D_{city} =|x1-x2|+|y1-y2|. \label{eq:city}
\end{equation}
The distance weight was set to squared inverse, where the weight is $1/distance^2$. 


\subsection{Performance Assessment}
Sensitivity, specificity, accuracy, and F1 score are the most often used performance measures in signal processing to evaluate the performance of an algorithm. The following equations are used to describe these metrics:
\begin{equation}
Sensitivity = \frac{TP}{TP + FN} 
\end{equation} 

\begin{equation}
Specificity = \frac{TN}{TN + FP} 
\end{equation} 

\begin{equation}
Accuracy = \frac{TP+TN}{TN + FP + TP + FN} 
\end{equation} 

\begin{equation}
Precision = \frac{TP}{TP + FP} 
\end{equation}

\begin{equation}
F1 \ Score = \frac{2*Precision*Sensitivity}{Precision + Sensitivity} 
\end{equation}

\noindent where $TP$ is the number of seizure periods that have been detected by both a human expert and the algorithm, and $FN$ is the number of seizure periods that have been determined by a human expert but have not been discovered by the algorithm. The number of nonseizure periods detected by both a human expert and the algorithm is represented by $TN$, and $FP$ defines the number of nonseizure periods that the algorithm detected as seizure but were not recognised as such by a human expert \cite{alotaiby2014eeg}. 

\section{Experimental Results and Discussion}
When comparing the ensemble bagged tree method against k-NN, the bagged tree method clearly outperforms k-NN, not only in binary classification but in all focal-generalised-nonseizure classification and multi-classification models. From the results presented in Table \ref{table:results_binl}, Table \ref{table:results_foc_gen}, and \ref{table:results_mult}, it is clear too that our novel channel selection outperforms ICA, and PCA.

\begin{table}[t]
\caption{Performance metrics for scenario (1): Binary Classification using the novel channel selection method, ICA, and PCA.} \label{table:results_binl}
\begin{tabular}{lllll} 
\hline
\multicolumn{1}{l|}{Method}      & Sensitivity & Specificity & Accuracy & F1-Score \\ \hline
\multicolumn{5}{c}{Novel Channel Selection}                                          \\ \hline
\multicolumn{1}{l|}{Bagged Tree} &\textbf{ 0.75   }     & \textbf{0.75}        & \textbf{0.82 }    & \textbf{0.73}     \\
\multicolumn{1}{l|}{k-NN}        & 0.73        & 0.73      & 0.77     & 0.68     \\ \hline
\multicolumn{5}{c}{ICA}                       \\ \hline
\multicolumn{1}{l|}{Bagged Tree} & 0.57        & 0.57        & 0.78     & 0.58     \\
\multicolumn{1}{l|}{k-NN}        & 0.56        & 0.56        & 0.66     & 0.54     \\ \hline
\multicolumn{5}{c}{PCA}                                           \\ \hline
\multicolumn{1}{l|}{Bagged Tree} & 0.62        & 0.62        & 0.74     & 0.61 \\
\multicolumn{1}{l|}{k-NN}        & 0.55        & 0.55        & 0.55     & 0.49     \\ \hline
\end{tabular}
\end{table}

\begin{figure}[h]
\begin{center}
\includegraphics[width=95mm,trim={1cm 3cm 2cm 1cm},clip]{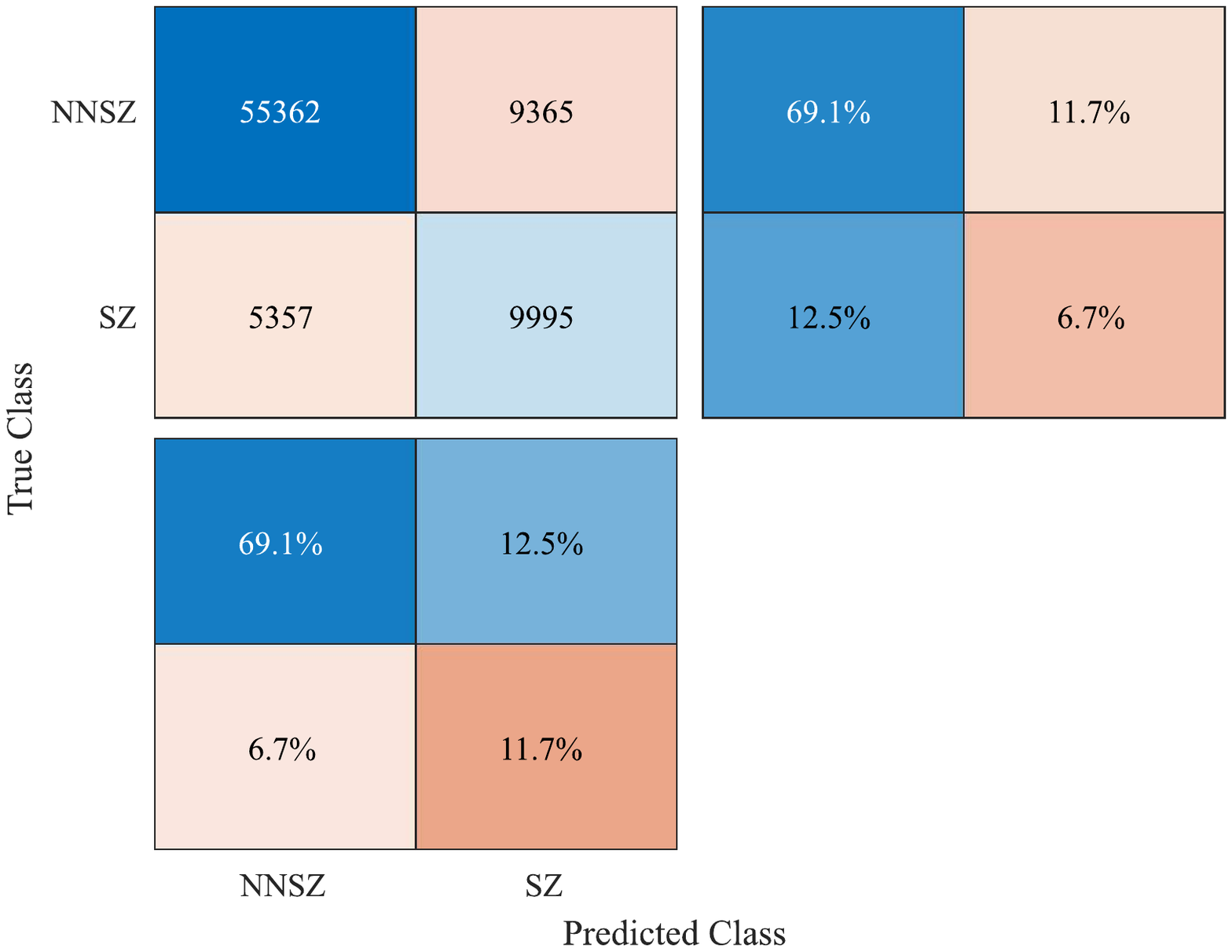}
\end{center}
\vspace*{-50mm}
\caption{Confusion Matrix depicting seizure (SZ) vs nonseizure (NNSZ) results from novel channel selection algorithm using bagged tree classifier.}
\label{fig:bin_novel_bag}
\end{figure}

Figure \ref{fig:bin_novel_bag} demonstrates a confusion chart of these results, where the classifier is able to accurately detect 0.86 of nonseizure periods and 0.65 of seizure periods, with an overall accuracy of 0.82, the highest results from this investigation. Whereas the poorest accuracy and F1 score are from using PCA with k-NN.

\begin{table}[t]
\caption{Performance metrics for scenario (2): Focal-generalised-nonseizure Classification using the novel channel selection method, ICA, and PCA.} \label{table:results_foc_gen}
\begin{tabular}{lllll}
\hline
\multicolumn{1}{l|}{Method}      & Sensitivity & Specificity & Accuracy & F1-Score \\ \hline
\multicolumn{5}{c}{Novel Channel Selection}                                          \\ \hline
\multicolumn{1}{l|}{Bagged Tree} & \textbf{0.59} & 0.77        & \textbf{0.68}    & \textbf{0.61 }    \\
\multicolumn{1}{l|}{k-NN}        & 0.64        & \textbf{0.79}      & 0.51     & 0.47     \\ \hline
\multicolumn{5}{c}{ICA}                       \\ \hline
\multicolumn{1}{l|}{Bagged Tree} & 0.47        & 0.74        & 0.40     & 0.32     \\
\multicolumn{1}{l|}{k-NN}        & 0.37        & 0.69        & 0.31     & 0.25     \\ \hline
\multicolumn{5}{c}{PCA}                                           \\ \hline
\multicolumn{1}{l|}{Bagged Tree} & 0.50        & 0.73        & 0.35     & 0.30 \\
\multicolumn{1}{l|}{k-NN}        & 0.42        & 0.69        & 0.31     & 0.25     \\ \hline
\end{tabular}
\end{table}

\begin{figure}[h]
\begin{center}
\includegraphics[width=9cm,trim={1cm 3cm 2cm 1cm},clip]{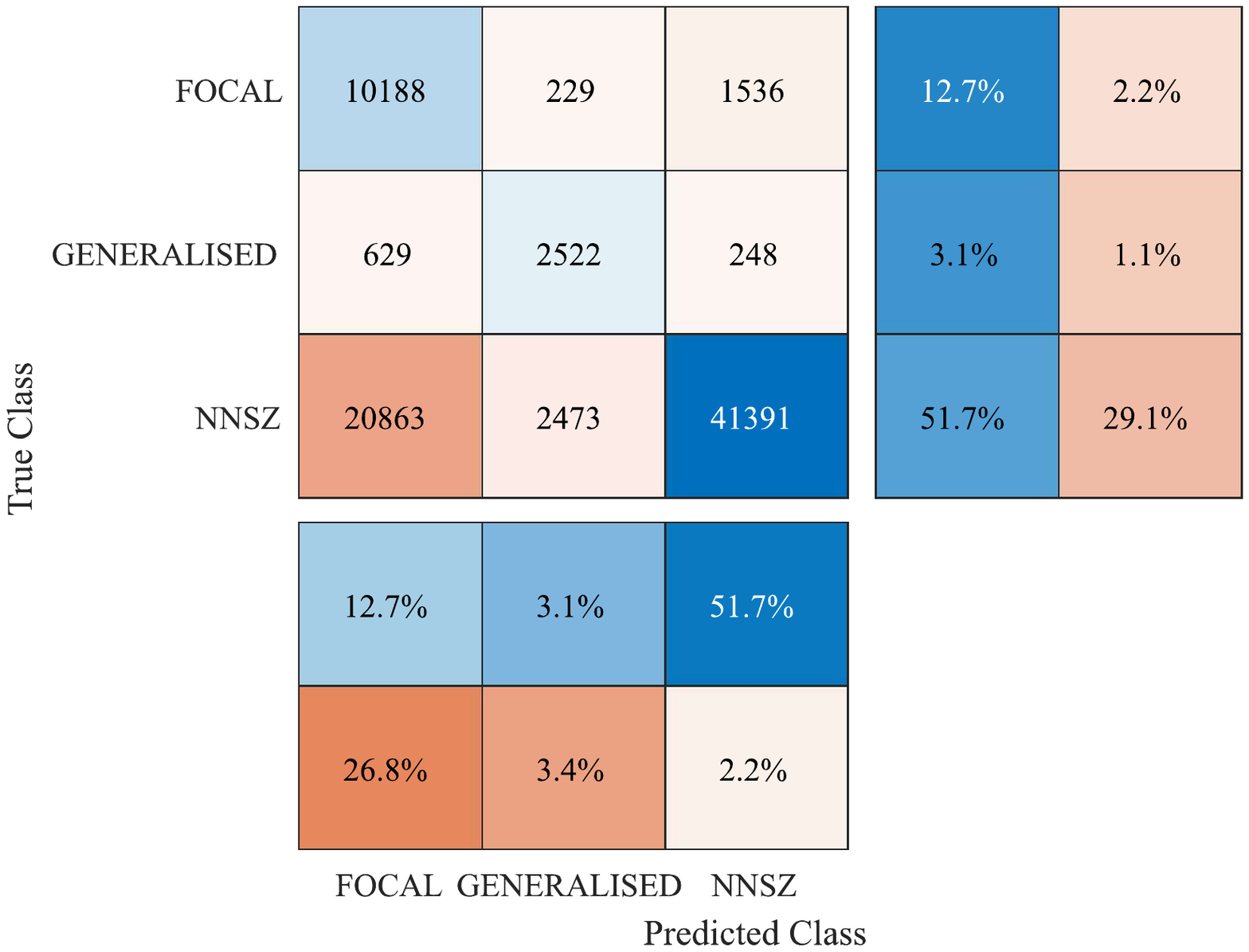}
\end{center}
\vspace*{-50mm}
\caption{Confusion Matrix depicting focal seizure, generalised seizure vs nonseizure results from novel channel selection algorithm using bagged tree classifier.}
\label{fig:foc_gen}
\end{figure}

When considering the scenario for focal-generalised-nonseizure classification as per Figure \ref{fig:foc_gen}, and Table \ref{table:results_foc_gen}, we see yet again the superiority of using the combination of the novel channel selection method and bagged tree classification. In this setting however, it seems that using the combination of ICA and k-NN, we get the poorest results. It is important to note that there are more cases of nonseizure periods being incorrectly classified as focal seizures as per Figure \ref{fig:foc_gen}. 

\begin{table}[t]
\caption{Performance metrics for scenario (3): Multi-classification using the novel channel selection method, ICA, and PCA.} \label{table:results_mult}
\begin{tabular}{lllll}
\hline
\multicolumn{1}{l|}{Method}      & Sensitivity & Specificity & Accuracy & F1-Score \\ \hline
\multicolumn{5}{c}{Novel Channel Selection}                                          \\ \hline
\multicolumn{1}{l|}{Bagged Tree} &\textbf{ 0.55} & \textbf{0.93}        & \textbf{0.69}    & \textbf{0.51}     \\
\multicolumn{1}{l|}{k-NN}        & 0.46        & 0.91      & 0.53     & 0.41     \\ \hline
\multicolumn{5}{c}{ICA}                       \\ \hline
\multicolumn{1}{l|}{Bagged Tree} & 0.26        & 0.89        & 0.47     & 0.47     \\
\multicolumn{1}{l|}{k-NN}        & 0.21        & 0.87        & 0.33     & 0.12     \\ \hline
\multicolumn{5}{c}{PCA}                                           \\ \hline
\multicolumn{1}{l|}{Bagged Tree} & 0.27        & 0.88        & 0.45     & 0.45 \\
\multicolumn{1}{l|}{k-NN}        & 0.25        & 0.87        & 0.32    & 0.32     \\ \hline
\end{tabular}
\end{table}

\begin{figure}[h]
\begin{center}
\includegraphics[width=9cm,trim={1cm 3cm 2cm 1cm},clip]{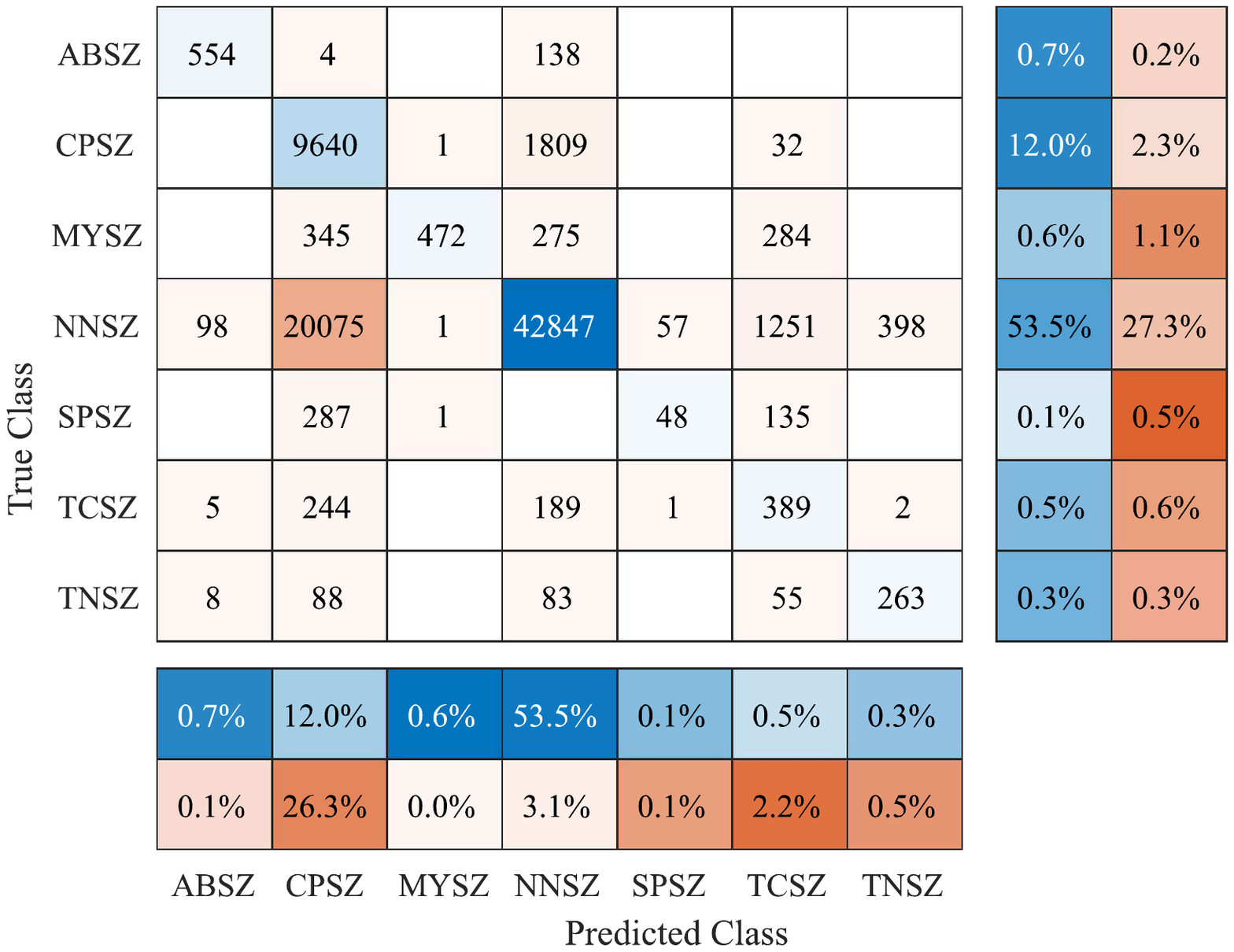}
\end{center}
\vspace*{-50mm}
\caption{Confusion Matrix depicting focal seizure, generalised seizure vs nonseizure results from novel channel selection algorithm using bagged tree classifier, where NNSZ is nonseizure period, SPSZ is a simple-partial seizure, CPSZ is a complex-partial seizure, MYSZ is a myoclonic seizure, ABSZ is an absence seizure, TNSZ is a tonic seizure, and TCSZ is a tonic-clonic seizure.}
\label{fig:mult_fb_bag}
\end{figure}

In the setting of multi-classification, overall accuracy has actually improved when comparing with focal-generalised-nonseizure classification. The novel channel selection algorithm using bagged tree classifier still achieved the best results. In this case, the poorest results are again gathered using ICA with k-NN. Interestingly from Figure \ref{fig:mult_fb_bag}, it can be observed to a greater extend than in Figure \ref{fig:foc_gen} that a relatively high quantity of nonseizure periods are being incorrectly identified. The confusion chart allows us to establish that the classifier incorrectly identifies some nonseizure cases as complex partial seizures.

\section{Conclusions}
The main challenge that exists today is obtaining an accurate automated seizure classification model that can differentiate between various seizure types to overcome the clinical burden of manual EEG analysis. This paper has presented the application of using our novel channel selection method based on frequency information dependent to specific regions of a normal brain. From the results presented, it can be noted that this method does successfully isolate the information found in an abnormal brain during seizure occurrences. It is evident that using this channel selection method combined with the bagged tree classification model outperforms other commonly used methods for seizure detection, with the highest accuracy at 0.82. From this experiment, it has also been discovered that there are difficulties with classification models when differentiating nonseizure periods from complex partial seizures. Therefore, future work will involve researching possible methods to isolate nonseizure periods from complex partial seizures. Furthermore, this work has proved promising in terms of multi-classification of the various seizure types. Future work can focus on improving this current methodology, with the possibility of moving more into deep learning methods to classify the various seizure types. Our novel channel selection with long short-term memory or various deep learning may provide better results, as there has been promising results recently using long short-term memory in seizure classification \cite{eeg}. It can also be noted that results gathered from scenario (3) are not much different from scenario (2), therefore we will redact scenario (2) from any future work.

\end{document}